\documentclass[epjCONF,columns]{svjour} 

\newcommand{\ppbar}  {\ensuremath{p\bar{p}} }
\newcommand{\ttbar}  {\ensuremath{t\bar{t}} }

\newcommand{\mtop}    {\ensuremath{\mathrm{M}_{\mbox{top}}} }
\newcommand{\wtop}    {\ensuremath{\Gamma_{\mbox{top}}} }
\newcommand{\dm}    {\ensuremath{\mathrm{\delta M}_{\mbox{top}}} }

\newcommand{\djes}   {\ensuremath{\Delta_{\mathrm{JES}}} }

\newcommand{\genunit}[2]{\ensuremath{#1~\mathrm{#2}} }

\newcommand{\gev}[1]    {\genunit{#1}{GeV}}

\newcommand{\gevcc}[1]  {\ensuremath{#1~\mathrm{GeV}/c^{2}}}

\newcommand{\invfb}[1]  {\ensuremath{#1~\mathrm{fb}^{-1}}}

\newcommand{\gevccnoarg}{\ensuremath{\mathrm{GeV}/c^{2}} }

\def\be{\begin{equation}}
\def\ee{\end{equation}}
\def\bea{\begin{eqnarray}}
\def\eea{\end{eqnarray}}

\usepackage{graphics}
\usepackage[varg]{txfonts} 
\usepackage[latin1]{inputenc}
\session-title{2011 Hadron collider Physics symposium (HCP-2011)}
\begin{document}
\title{Top quark mass and property measurements at Tevatron}
\author{Hyun Su Lee\thanks{\email{hslee@fnal.gov}} on behalf of the CDF and D0 collaborations }
\institute{Korea University, Seoul 136-713, Korea}
\abstract{
The top quark, discovered in 1995 at the Fermilab Tevatron collider from CDF and D0 experiments, remains by far the most interesting particle to test standard model because of its large mass and unique properties. Having data collected about 10 fb$^{-1}$ of integrated luminosity of $p\bar{p}$ collision, both experiments have been studied the top quark in all the possible directions. In this article, we present the recent measurements of the top quark properties from Tevatron including the mass, width, spin correlation, and $W$ boson helicity using $t\bar{t}$ signature. 
} 
\maketitle
\section{Introduction}
\label{intro}
The top quark, observed by both the CDF and D0 experiments in 1995~\cite{top_discovery}, is by far the heaviest known elementary particle and its mass is almost 40~times heavier than its isospin partner, the bottom~($b$) quark~\cite{pdg}. Due to the heavy mass, the top quark plays an important role in electroweak radiative corrections relating the top quark mass~(\mtop) and the $W$ boson mass to the mass of the predicted Higgs boson~\cite{ewfit1,ewfit2}. The lifetime of the top quark is about 20 times shorter than the timescale for strong interactions, and therefore it does not form hadrons, giving us a unique opportunity to study a ``bare'' quark. 

Top quarks at the Tevatron are predominantly produced in pairs, and decay almost always to a $W$ boson and a $b$ quark in the standard model (SM). The topology of \ttbar events depends on the different decay of the two $W$ bosons. In the dilepton channel, each $W$ boson decay to charged lepton~(electron and muon) and neutrino. Events in this channel thus contain two leptons, two $b$-quark jets, and two undetected neutrinos. Because of the presence of two leptons, this channel has the lowest background. However the dilepton channel has the smallest branching fraction. In the all-jets channel, each $W$ boson decays to two jets so that this channel contains two $b$ quark jets and four light quark jets. This channel has the largest branching fraction but also the largest background from QCD multijet production. The lepton+jets channel has one $W$ boson decaying leptonically and the other hadronically so that we have one charged lepton, two $b$-quark jets, two light quark jets, and one undetected neutrino. Because of the relatively large branching fraction with manageable backgrounds, lepton+jets channel is considered as the ``golden channel'' in the top quark studies. By this reason, the most results presented here use the lepton+jets final state.

\section{Measurements}
\label{sec:1}
\subsection{Top quark mass}
The mass of the top quark is very important to estimate the SM Higgs boson because precise top and $W$ boson masses measurements can predict the mass of the Higgs boson either SM or beyond SM. Since the discovery of the top quark, both the CDF and D0 experiments have been improving the precision of the \mtop measurement~\cite{topmasscomb}. 

For the \mtop measurements, two primary techniques have been established. The template method~(TM) uses the distributions of variables~(templates) which are strongly correlated with the top quark mass and jet energy scale~(JES). In the building of a probability, only a few variables (usually less than two) are used, for instance reconstructed top quark mass and dijet mass of hardronic decay $W$ boson in the lepton+jets channel.  
The Matrix Element Method~(ME) uses event's probability to be a combinates signals and background. ME exploit all the information in the event by using a leading order matrix element calculation convoluted with parton distribution function and transfer functions~(TFs) making connection between detector response and parton level particle. Because we can use all the information of $\ttbar$ production and decay in principle, ME usually provide better precision of \mtop than TM. Both techniques employ likelihood to compare data to the modeling of signals and background to extract \mtop. 

\begin{figure}
\begin{center}
\resizebox{0.8\columnwidth}{!}{%
\includegraphics{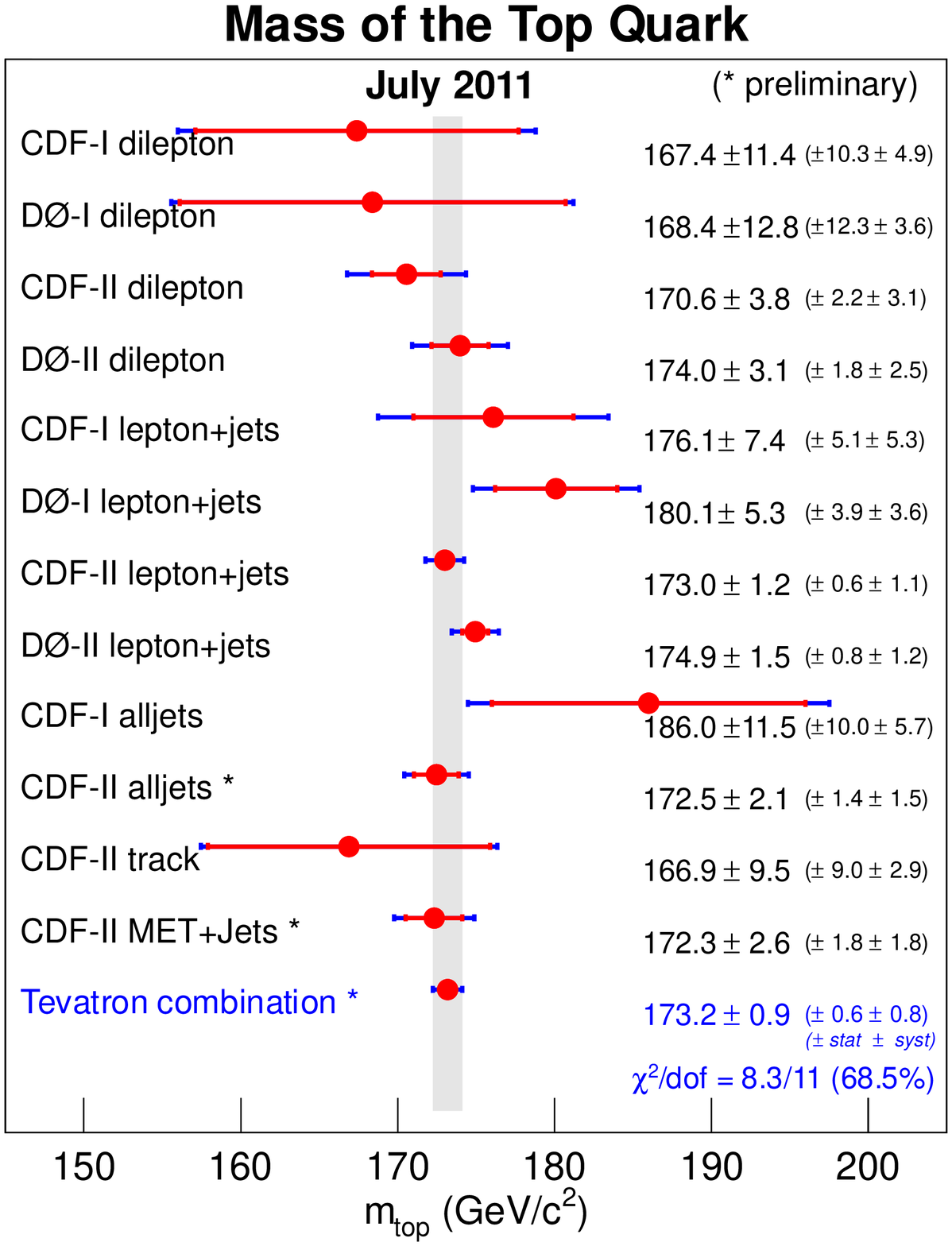}}
\resizebox{0.8\columnwidth}{!}{%
\includegraphics{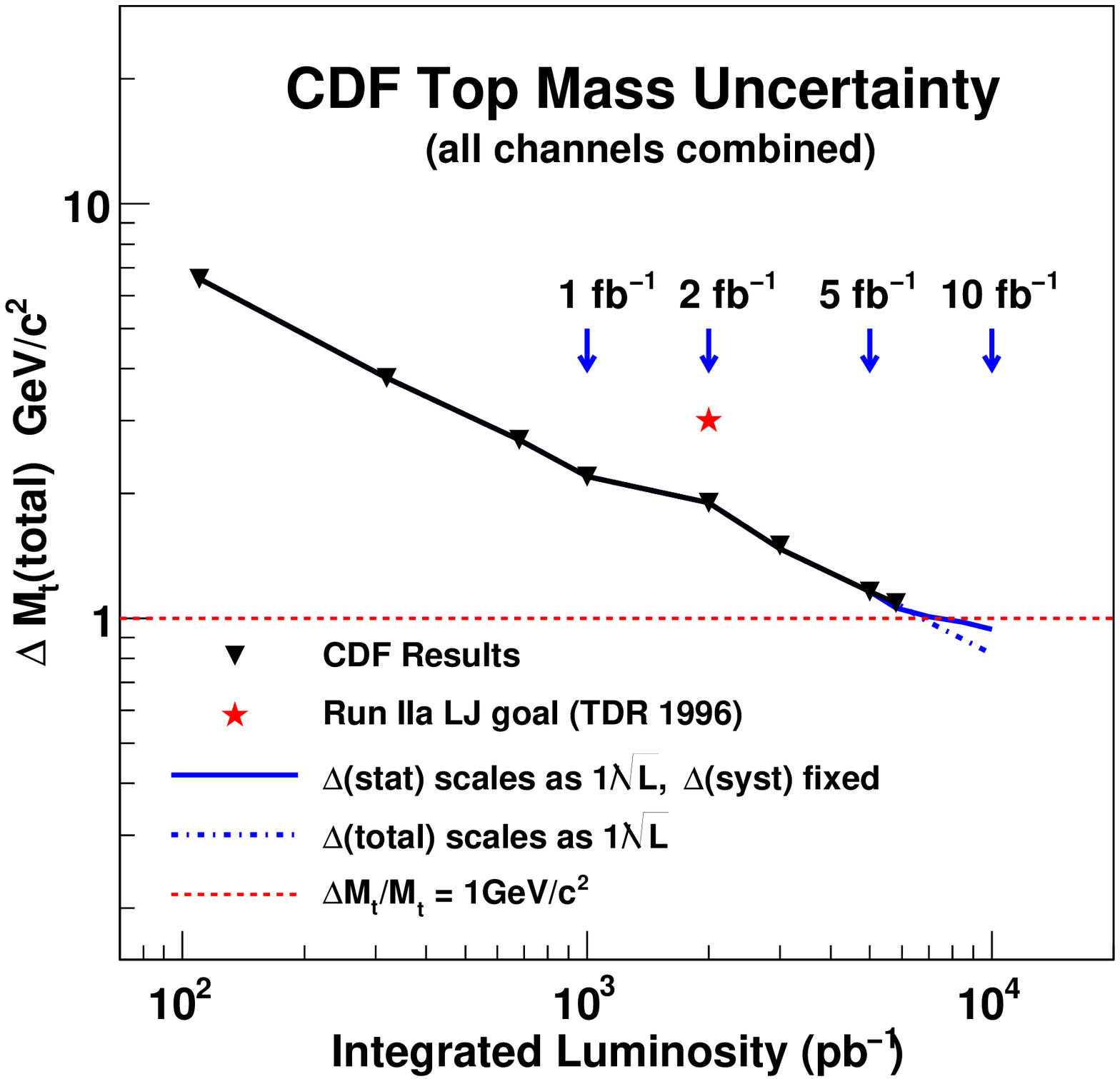}}
\end{center}
\caption{ Up: Summary of the Tevatron top quark mass measurements and its combination. Down: CDF prediction of \mtop precision by scaling using increased luminosity~(solid line) and plus possible improvement~(dashed line).
}
\label{ref:mtop}
\end{figure}

CDF and D0 experiments have performed the \mtop measurements in the various final states with different techniques. In the lepton+jets and all-jets channels the uncertainty from JES can be reduced by using the reconstructed dijet mass from hadronically decaying $W$ boson with {\it in situ} calibration of JES. To date the most precise measurement has been performed by CDF collaboration using lepton+jets channel with ME. We found \mtop = \gevcc{173.0 \pm 1.2} with \invfb{5.6} of the data~\cite{cdfljmass}. D0 carried out the most precise \mtop measurement in the dilepton channel using TM. We built templates of the reconstructed top quark mass distributions and extract \mtop = \gevcc{173.3 \pm 3.2} using \invfb{5.3} data~\cite{d0dilmass}. CDF have interesting measurements using all hadronic final state with and without large missing energy. The large missing energy and jets final state is mostly originated from lepton+jets events but, lepton is not reconstructed by hadronically decaying $\tau$ lepton or going outside of detector coverage. These events have not been used for \mtop measurement and CDF firstly start use it to measure \mtop = \gevcc{172.3 \pm 2.6} with \invfb{5.7} of data~\cite{cdfmetjet}. The jets without missing energy is coming from all jet channel and CDF collaboration measure \mtop = \gevcc{172.5 \pm 2.0} using \invfb{5.8} of data~\cite{cdfalljet}
Figure~\ref{ref:mtop} (left) shows the summary of the \mtop measurements and the combination of the Tevatron~\cite{topmasscomb} which is \mtop = \gevcc{173.2 \pm 0.9}.  The precision, $\Delta \mtop/\mtop \sim 0.6$\%, is already surpassed the prediction of RunII experiments and reach to less than \gevcc{1}. We predict to reach less than \gevcc{1} precision by end of RunII with over \invfb{10} data in each experiment as shown in Fig.~\ref{ref:mtop} (right). 

The precision determination of \mtop allows us to measure the mass difference between top quark and anti-top quark to a few GeV. In the CPT theorem, which is fundamental to any local Lorentz-invariant quantum field theory, the quark mass should be same as the mass of its anti-quark partner. Despite the fact that no violations have ever been observed in the meson and baryon sectors, it is important to test CPT violation in all possible sectors such as quarks and high mass particles. 

D0 collaboration has direct measurements of the top quark and the anti-top quark mass difference~(\dm) in the lepton+jets channel using the ME. 
In the matrix element calculation, one assumes SM-like \ttbar production and decay, where identical particle and antiparticle masses are assumed for $b$ quarks and W bosons but not for top quarks. Using \invfb{3.6} of \ppbar collision data, we measure \dm~=~0.8~$\pm$~1.9~\gevccnoarg~\cite{d0cpt}. 
CDF collaboration has a measurement using the TM. We reconstruct the mass difference using modified kinematic fitter allowing mass difference between hadronic top quark and leptonic top quark. Using \invfb{5.6} of \ppbar collisions, we measure $\dm = -3.3 \pm 1.7 \gevccnoarg$~\cite{cdfcpt}. It is consistent with CPT symmetry at a 2 standard deviation level. 

\subsection{Top quark width}
Because of the short lifetime, a direct determination of the top quark lifetime is extremely hard. However, we can calculate it from the decay width. CDF collaboration has a direct measurement of the top quark width~(\wtop) using \invfb{4.3} of \ppbar collision. The \mtop and the mass of W boson that decays hadronically are reconstructed for each event and compared with templates of different \wtop and deviations from nominal jet energy scale~(\djes) to perform a simultaneous fit for both parameters, where \djes is used for the {\it in situ} calibration of the jet energy scale. By applying a Feldman-Cousins approach, we establish an upper limit at 95~\% confidence level of $\wtop < \gev{7.6}$ and a two-sided 68~\% CL interval of $\gev{0.3} < \wtop < \gev{4.4}$~\cite{cdfwidth}. D0 collaboration has an indirect determination of \wtop using single top t-channel cross section and $(t\rightarrow Wb)/(t\rightarrow Wq)$ fraction measurements. The \wtop is calculated with quantum mechanical relation, $\wtop = \frac{\sigma(t-ch)}{Br(t\rightarrow bW)}\cdot \frac{Br(t\rightarrow bW)_{\mathrm{SM}}}{\sigma(t-ch)_{\mathrm{SM}}}$. The result, $\wtop = \gev{1.99^{+0.65}_{-0.55}}$, is the most precise determination of the top quark width using experimental data sample and consistent with the SM~\cite{d0width}.

\subsection{Spin Correlation}

The \ttbar spin correlation is predicted by the SM and a potentially sensitive discriminant of new physics coupled to the top quark. The spin state is observable in angular correlations among the quark decay products. In the dilepton channel, we used the angular correlation between two leptons and measured consistent results with the SM from both CDF~\cite{cdfdilspin} and D0~\cite{d0dilspin} collaborations. CDF collaboration  has a measurement, consistent with the SM, using lepton+jets channel with information of the correlation between lepton and down-type quark~\cite{cdfljspin}.

\begin{figure}
\begin{center}
\resizebox{0.8\columnwidth}{!}{%
\includegraphics{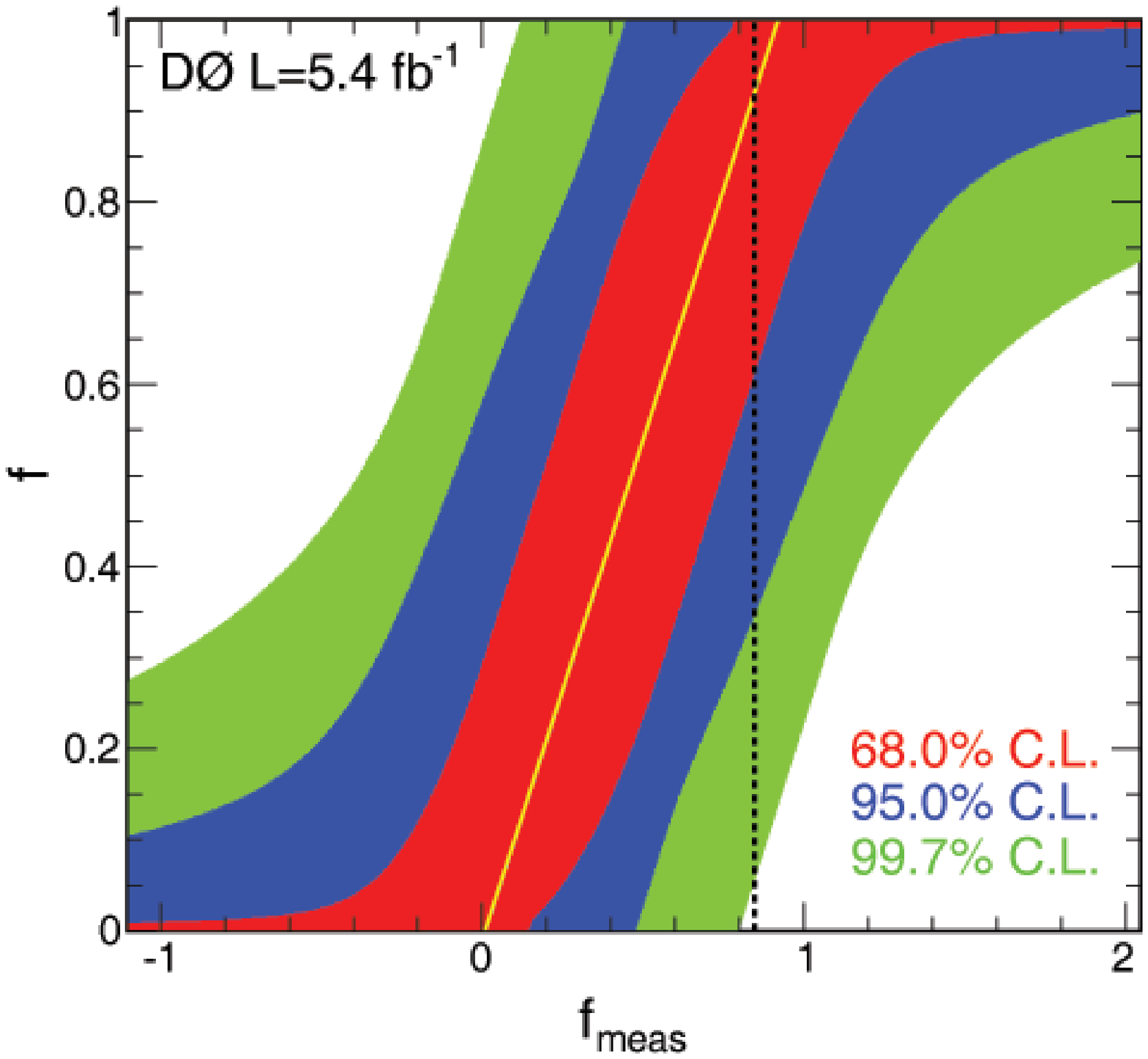}}
\resizebox{0.8\columnwidth}{!}{%
\includegraphics{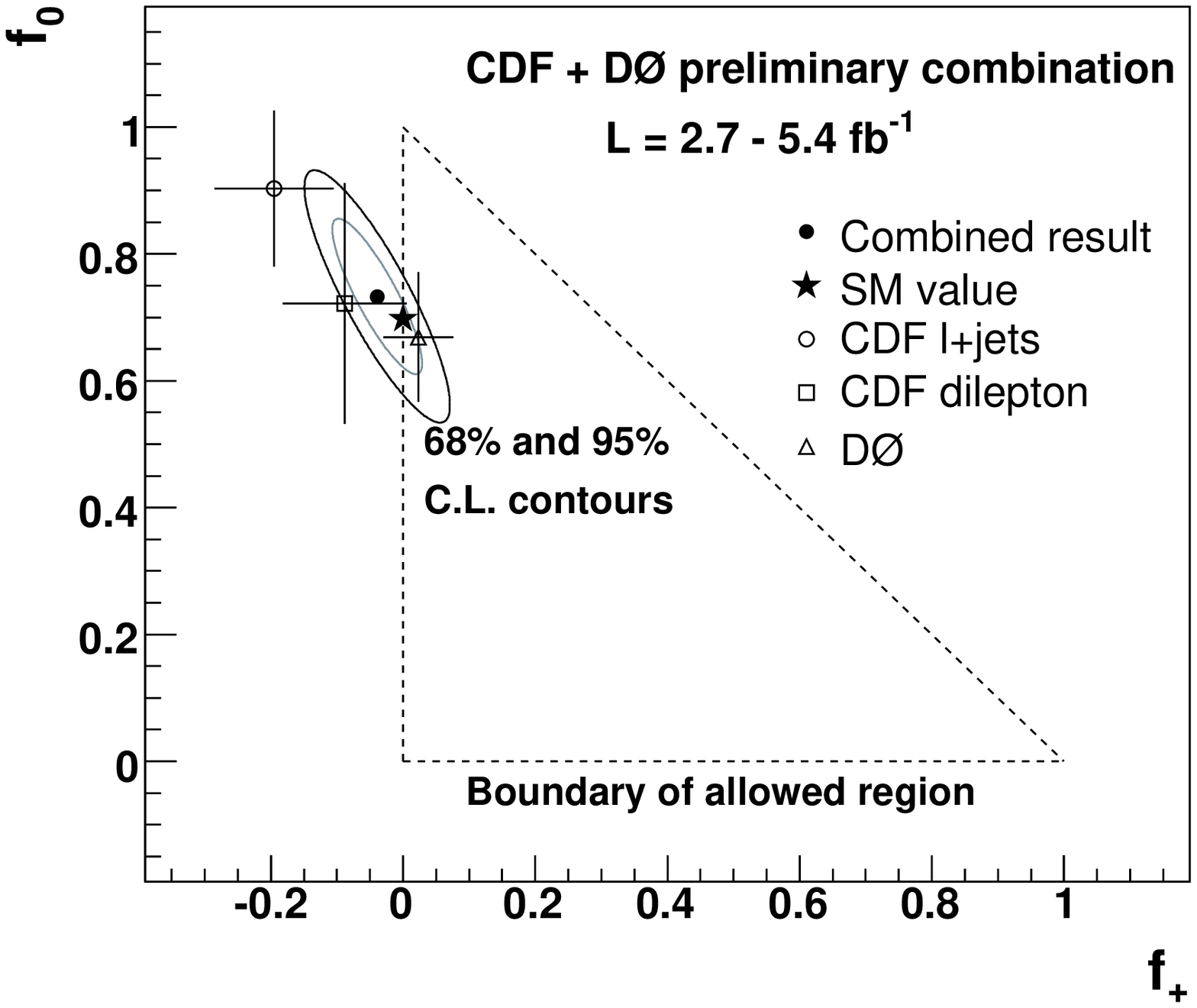}}
\end{center}
\caption{ Up: Bands of spin correlation coefficient~(f) as a function of measured value for the combined dilepton and lepton+jets measurement from D0: The vertical dotted line indicated the measured value of 0.85. Down: Combined tevatron result on W boson helicity measurement with 1 and 2 standard deviation contours are compared with the SM prediction. 
}
\label{fig:spinWhe}
\end{figure}

Although these results are consistent with the SM, we are not demonstrating the spin correlation because of large statistical uncertainty. There are a couple of development of new observables~\cite{spin_the1}~\cite{spin_the2} to improve spin correlation measurement. Especially, Ref.~\cite{spin_the2} suggest to use a likelihood ratio R~($=\frac{P(H=c)}{P(H=u)+P(H=c)}$) calculated from leading order matrix element probability in case of spin correlation~($P(H=c)$) and no spin correlation~($P(H=u)$). D0 collaboration employed this technique for dilepton channel~\cite{d0spin_prl} and lepton+jets channel~\cite{d0spin_combo}. Combined fit with two channel using this technique shows spin correlation coefficient~(f) greater than 0.05 at three standard deviation as one can see in Fig.~\ref{fig:spinWhe} (left). This  means that we reject no spin correlation~(f=0) case with 99.7\% CL. This is the first evidence for the presence of spin correlation with a significance of more than three standard deviations~\cite{d0spin_combo}. 

\subsection{W boson helicity}
The SM predicts that the top quark decays almost entirely to a $W$ boson
and a bottom quark, and that the $Wtb$ vertex is a V-A charged weak
current interaction. A consequence of this is that approximately 70\% of the top quark 
decay longitudinally, 30\% of the top quarks have left handed polarization~(f$_0$ = 70\%, f$_{-}$=30\%, f$_{+}$=0\%)~\cite{wpred}. Any new particles involved in the same
decay topologies and non-standard coupling could create a different
mixture of polarized $W$ bosons. Therefore, a measurement of this
fraction is a test of the V-A nature of the $Wtb$ vertex. 
D0 collaboration uses both lepton+jets and dilepton channel simultaneously with \invfb{4.3} data and extracts
 f$_{+} = 0.02 \pm 0.05$ and f$_{0} = 0.67 \pm 0.10$ with the simultaneous fit of the two variables~\cite{d0whel}. 
CDF collaboration has results in both lepton+jets~\cite{cdfwhellj} and dilepton channels~\cite{cdfwheldil}. All three different measurement are combined to get  f$_{+} = -0.013 \pm 0.035$ and f$_{0} = 0.685 \pm 0.057$ which is consistent with the SM prediction as one can see in Fig.~\ref{fig:spinWhe} (right). 

\subsection{Forward backward charge assymetry~($A_{fb}$)}
At the leading order, \ttbar production at the SM is symmetric. However, at the higher order, interference between diagrams yields positive charge asymmetry, for instance approximately 5-7\% asymmetry at the next leading order~\cite{afb_theory}. If heavy new particle, such as Z prime~(Z boson like heavy particle) and axigluon, mediate \ttbar production, this asymmetry can be much enhanced.

\begin{figure}
\begin{center}
\resizebox{0.8\columnwidth}{!}{%
\includegraphics{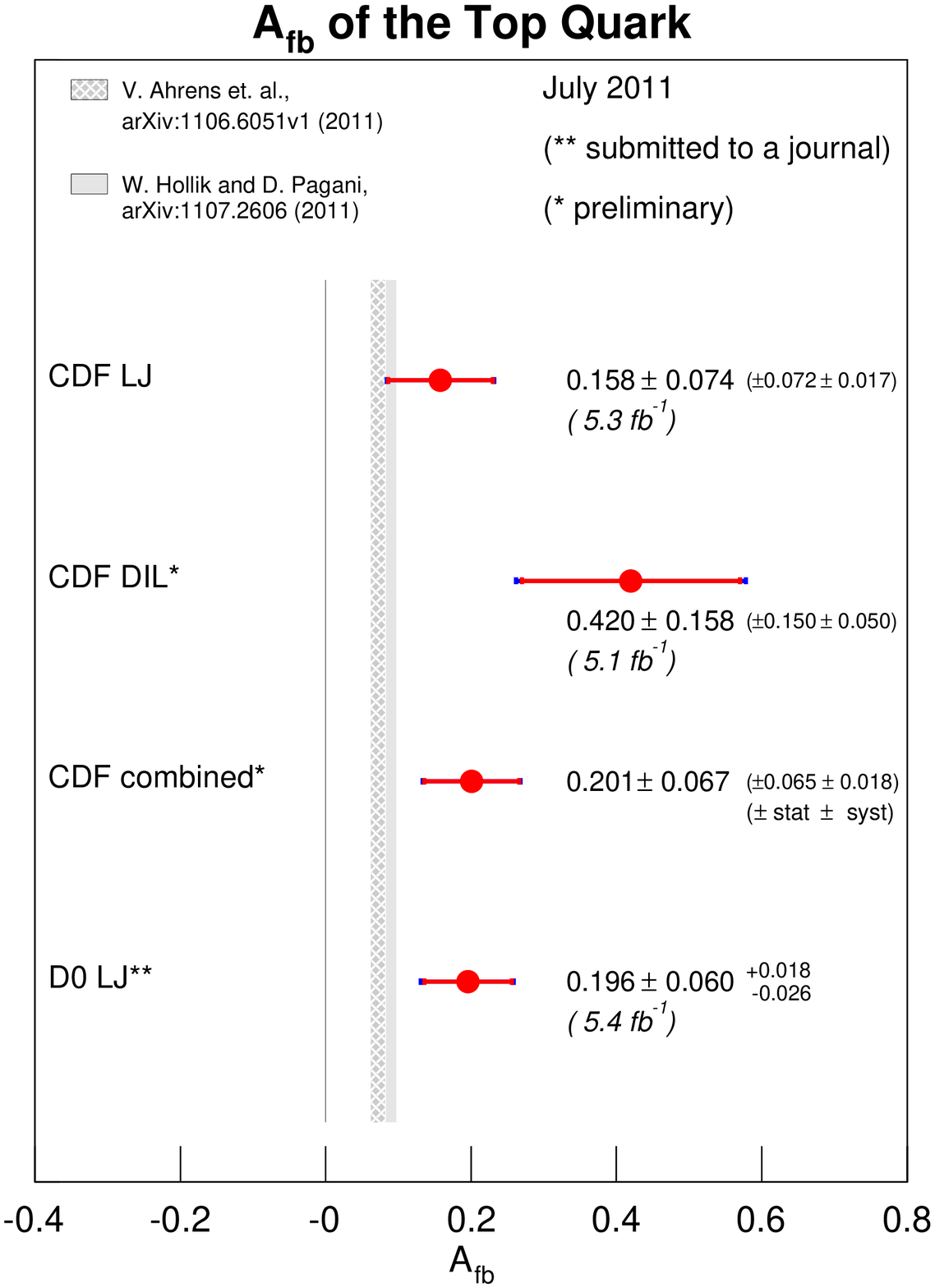}}
\resizebox{0.8\columnwidth}{!}{%
\includegraphics{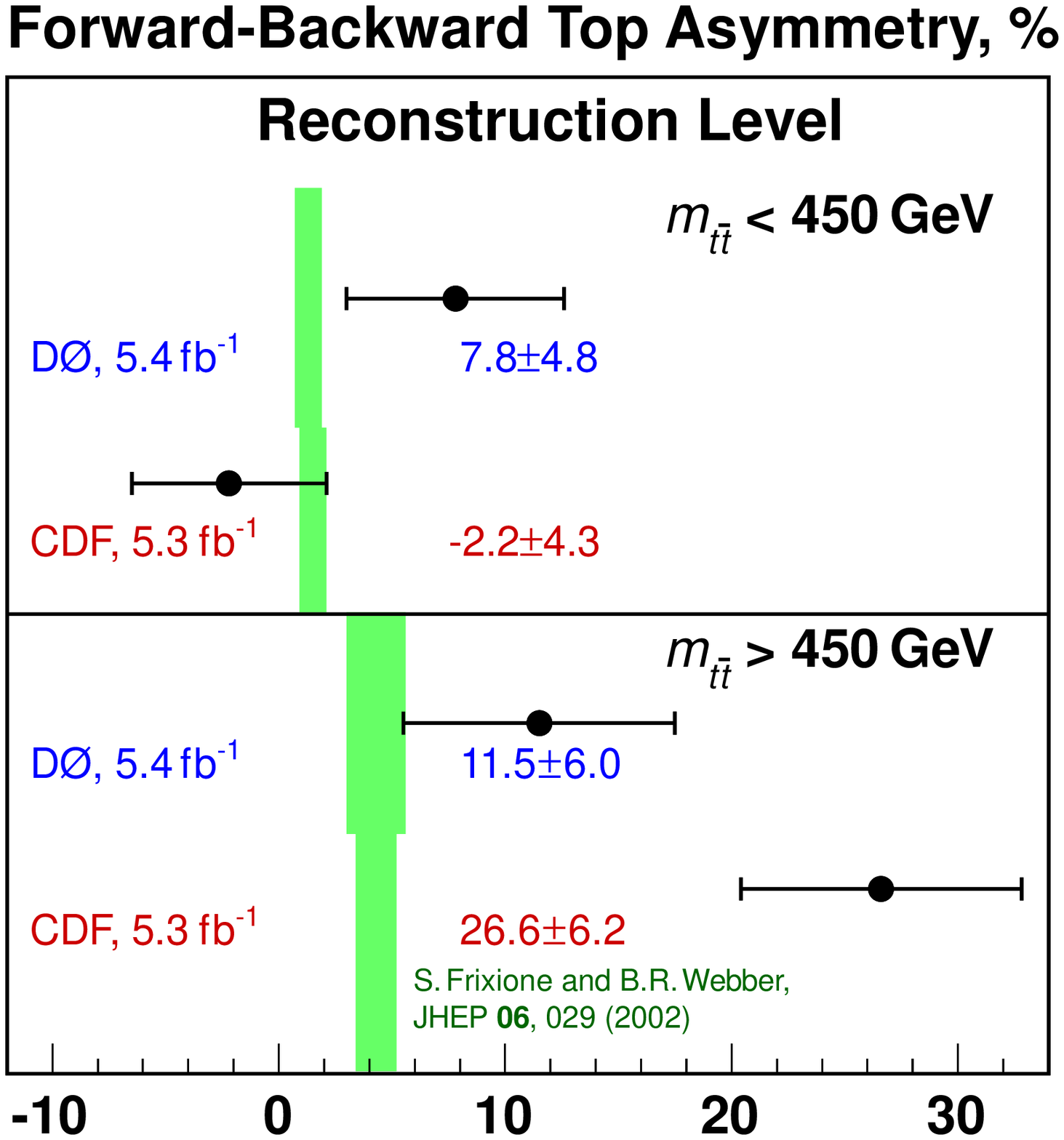}}
\end{center}
\caption{ Up: Summary of the $A_{fb}$ measurements is shown. Down: $A_{fb}$ measurements of $m_{\ttbar}<450 GeV$~(up) and $m_{\ttbar}>450 GeV$~(down) for CDF and D0 are compared with theory prediction~(green band).
}
\label{fig:afb}
\end{figure}

CDF collaboration measured an inclusive asymmetry of $A_{fb} = 0.158 \pm 0.074$ using \invfb{5.3} of data in the lepton+jets channel~\cite{cdfAfblj} and $A_{fb} = 0.420 \pm 0.158$ in the dilepton channel~\cite{cdfAfbdil}. The combined result of these two measurements shows $A_{fb} = 0.201 \pm 0.067$. D0 has a measurement of $A_{fb} = 0.196 \pm 0.060^{+0.018}_{-0.026}$ using \invfb{5.4} of lepton+jets channel data~\cite{d0Afblj}. All measurements are summarized in Fig.~\ref{fig:afb} (left) with selected theory prediction~\cite{afb_theory}. Uncertainties of all results are still dominated by statitical uncertainties. Comparing with theory predictions Tevatron measurements have approximately two standard deviation difference towards higher values of $A_{fb}$. 

Beside inclusive measurement, it is interesting to investigate the dependence of the asymmetry on various different regions such as the rapidity and the invariant mass of \ttbar~($m_{\ttbar}$). CDF and D0 investigated the $m_{\ttbar}$ dependence by measuring $A_{fb}$ for $m_{\ttbar}<450 GeV$ and $m_{\ttbar}>450 GeV$ regions as one can see in Fig.~\ref{fig:afb}~(right). While D0 data do not have significant dependence on $m_{\ttbar}$, CDF have clear dependence on $m_{\ttbar}$ resulting approximately three standard deviation from next leading order prediction for $m_{\ttbar}>450 GeV$. 

It is not yet clear source of inclusive asymmetry and different dependence on $m_{\ttbar}$ from two experiments. While all measurement have limited result caused by statistical uncertainty, the final measurement using over \invfb{10} data will be very interesting. 
\section{Conclusion}
The CDF and D0 collaborations have performed a robust set of analyses using many techniques and improvements to have better understanding the top quark nature. With Tevatron shutdown at last year, we have $>$10~fb$^{-1}$ of data acquired in each experiment which could be almost a double the data sample used in this report. An ultimate precision of approximately \mtop  less than \gevcc{1} in each experiment will be possible. The other top properties, which are mostly limited by statistics, have been significantly improved and we may have surprising results.

\section*{Acknowledgments}
I would like to thank for the CDF and D0 colleagues for their efforts to carry out these challenging physics analyses. I also thank for the conference organizers for a very rich week of physics. My travel was supported by 
the National Research Foundation of Korea Grant funded by the Korean Government [NRF-2011-35B-C00007].

\end{document}